\documentclass{article}

\usepackage{microtype}
\usepackage{graphicx}
\usepackage{subfigure}
\usepackage{booktabs} 

\usepackage{amsfonts}       
\usepackage{dsfont}       

\usepackage{amsmath}

\usepackage{algorithm}
\usepackage{algpseudocode}
\usepackage{listings}

\def\chapterautorefname~#1\null{%
  Chapter~#1\null
}
\def\sectionautorefname~#1\null{%
  Section~#1\null
}
\def\subsectionautorefname~#1\null{%
  Subsection~#1\null
}
\def\equationautorefname~#1\null{%
  (#1)\null
}
\def\algorithmautorefname~#1\null{%
  Algorithm~#1\null
}
\def\conjautorefname~#1\null{%
  Conjecture~#1\null
}

\renewcommand{\algorithmicrequire}{\textbf{Inputs:}}
\renewcommand{\algorithmicensure}{\textbf{Outputs:}}

\newcommand{\pr}{{\mathcal P}} 
\newcommand{\D}{{\mathcal D}} 
 
\newcommand{\x}{{\mathbf x}}

\newcommand{\y}{{\mathbf y}} 
\newcommand{\A}{{\mathbf A}}

\newcommand{\eg}{\textit{e.g.}\ }
\newcommand{\ie}{\textit{i.e.}\ }

\newcommand{\argmin}[1]{\arg\underset{#1}{\min}}

\usepackage{hyperref}

\usepackage[accepted]{icml2019}

\icmltitlerunning{}

\begin{document}

\twocolumn[
\icmltitle{How to do Physics-based Learning}



\icmlsetsymbol{equal}{*}

\begin{icmlauthorlist}
\icmlauthor{Michael Kellman}{berkeley}
\icmlauthor{Michael Lustig}{berkeley}
\icmlauthor{Laura Waller}{berkeley}
\end{icmlauthorlist}

\icmlaffiliation{berkeley}{Electrical Engineering and Computer Sciences, University of California, Berkeley, USA}

\icmlcorrespondingauthor{Michael Kellman}{kellman@berkeley.edu}

\icmlkeywords{Computational Imaging, Physics-based Learning, Tutorial}

\vskip 0.3in
]



\printAffiliationsAndNotice{} 

\section{Introduction}
Computational imaging systems (\eg tomographic systems, computational optics, magnetic resonance imaging) jointly design software and hardware to retrieve information which is not traditionally accessible. Generally, such systems are characterized by how the information is encoded (forward process) and decoded (inverse problem) from the measurements. Critical aspects of computational imaging systems, such as experimental design and image priors, can be optimized through deep networks formed by \textit{unrolling} the iterations of classical model-based reconstructions~\cite{Gregor:2010, sun2016deep, kellman2019physics}. Termed physics-based learning, this paradigm is able to incorporate knowledge we understand, such as the physics-based forward model and optimization methods, while relying on data-driven learning for parameters that we do not. A complete review the field can be found here~\cite{ongie2020deep}.

The goal of this tutorial is to explain step-by-step how to implement physics-based learning for the rapid prototyping of a computational imaging system. We provide a basic overview of physics-based learning, the construction of a physics-based network, and its reduction to practice. Specifically, we advocate exploiting the auto-differentiation functionality~\cite{Griewank2018evalder} twice, once to build a physics-based network and again to perform physics-based learning. Thus, the user need only implement the forward model process for their system, speeding up prototyping time. We provide an open-source Pytorch~\cite{paszke2017automatic} implementation~\footnote{\url{https://github.com/kellman/physics_based_learning}} of a physics-based network and training procedure for a generic sparse recovery problem (Sec.~\ref{sec:implementation}).

\section{Background}

Computational imaging systems are described by how sought information is encoded to and decoded from a set of measurements. The encoding of information, $\mathbf{x}$, into measurements, $\mathbf{y}$, is given by
\begin{align}
    \mathbf{y} = \mathcal{A}(\mathbf{x}) + \mathbf{n},
\end{align}
where $\mathcal{A}$ describes the forward model process that characterizes the formation of measurements and $\mathbf{n}$ is random system noise. The image reconstruction from a set of measurements, \ie decoding, can be structured using an inverse problem formulation, 
\begin{align}
    \x^\star = \argmin{\x} \ \underset{\D(\x;\y)}{\underbrace{\|\mathcal{A}(\x) - \y\|^2}} + \pr(\x),
\end{align}
\noindent where $\x$ is the sought information, $\D(\cdot)$ is the data consistency penalty (commonly $\ell_2$ distance between the measurements and the estimated measurements), and $\pr(\cdot)$ is the signal prior (\eg sparsity, total variation). This optimization problem often requires a non-linear and iterative solver. Proximal gradient descent can efficiently solve the optimization problem in the case of a non-linear signal prior~\cite{parikh2014proximal}. Methods such as Alternating Direction Method of Multipliers (ADMM)~\cite{boyd2011distributed} and Half-Quadratic Splitting (HQS)~\cite{geman1995nonlinear} can be efficient solvers when multiple constraints are placed on the image reconstruction and the forward model process is linear.

\begin{figure}[H]
    \centering
    \includegraphics[width=8.5cm, page=1]{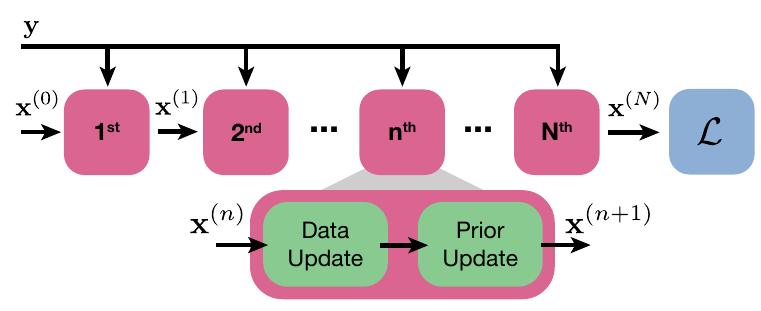}
    \caption{Unrolled Physics-based network composed of $N$ unrolled decoder iterations. Each layer is comprised of a data consistency update (\eg gradient update) and a prior update (\eg proximal update). The measurements, $\y$, and initialization, $\x^{(0)}$, serve as the network's inputs and the final estimate, $\x^{(N)}$, as the network's output. Finally, the final estimate is fed into some loss function, $\mathcal{L}$, for training.}
    \label{fig:illustration}
\end{figure}

The physics-based network (Fig.~\ref{fig:illustration}) is then formed by unrolling $N$ iterations of the optimization algorithm into the layers of a network. The inputs to the network are the measurements and initialization and the output is an estimate of the information after $N$ iterations of the optimizer.

\section{Implementation}
\label{sec:implementation}
For the purposes of demonstration, we implement physics-based learning on a compressed sensing problem -- under-determined sparse recovery from linear Gaussian random measurements. The sparse signals have normally-distributed amplitude values. The learnable parameters in the physics-based network will be the measurement matrix, the sparsity prior penalty, and the step size of the optimizer. Specifically, we solve the $\ell_1$ relaxation of the sparse recovery problem,
\begin{align}
    \x^\star = \argmin{\x} \ \|\A \x - \y\|^2 + \lambda \|\x\|_1 \text{,} \label{eq:sparse_recovery}
\end{align}
\noindent where $\lambda$ is the sparsity prior penalty that trades off the penalty of the prior with the data consistency term, and $\A \in \mathds{R}^{J \times K}$ is the linear measurement matrix.

We use the proximal gradient descent algorithm (Alg.~\ref{alg:pgd}) to solve the optimization problem (Eq.~\ref{eq:sparse_recovery}) and to form the architecture of the physics-based network.

\begin{algorithm}
    \caption{Proximal Gradient Descent}
    \label{alg:pgd}
    \textbf{Inputs} $\mathbf{x}^{(0)}$-initialization, $\alpha$-step size, $N$-number of iterations, $\y$-measurements \\
    \textbf{Output} $\mathbf{x}^{(N)}$-final estimate of image
    \begin{algorithmic}[1]
        \State $n \gets 0$
        \For{$n < N$}
        \State $\mathbf{z}^{(n)}  \gets \mathbf{x}^{(n)} - \alpha \nabla_\mathbf{x}\mathcal{D}(\mathbf{x}^{(n)};\mathbf{y})$ \label{alg:line:pgd_dc}
        \State $\mathbf{x}^{(n+1)} \gets \texttt{soft\_thr}_{\alpha \lambda}(\mathbf{z}^{(n)})$ \label{alg:line:pgd_prox}
        \State $n \gets n + 1$
        \EndFor
    \end{algorithmic}
\end{algorithm}

\subsection*{Physics-based network}

\begin{figure}
    \centering
    \includegraphics[width=8.5cm]{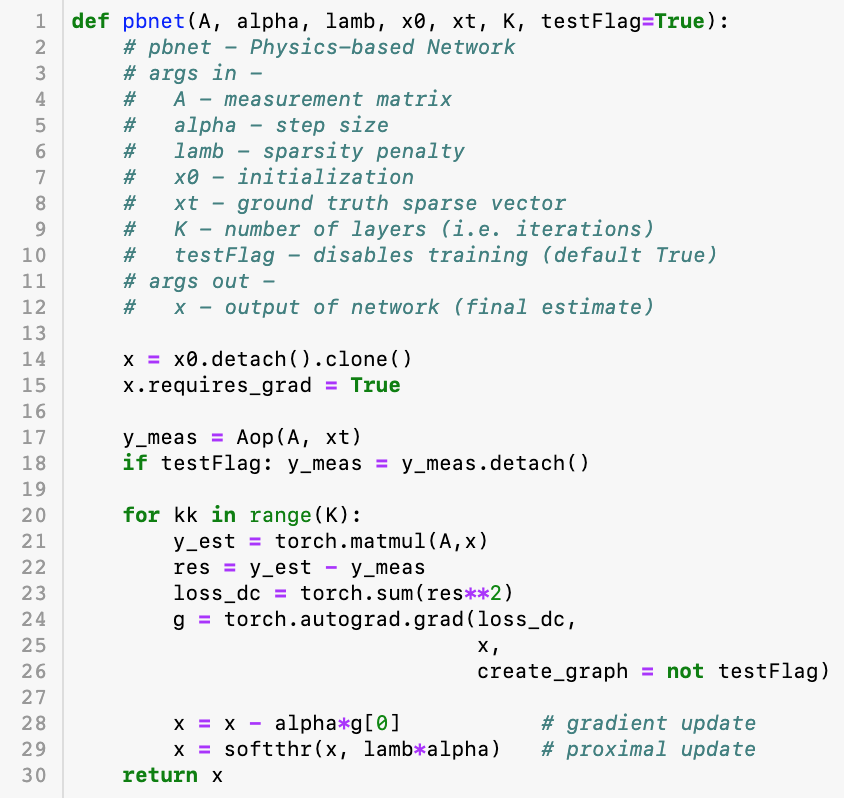}
    \caption{Implementation of a physics-based network for compressed sensing sparse recovery using automatic differentiation to compute gradients with respect to $\mathbf{x}$. The network is setup to learn the measurement matrix, step size, and sparsity penalty. Soft thresholding is used as the proximal operator.}
    \label{fig:pbnet}
\end{figure}

With some extra Pytorch specific flags and functionalities, the physics-based network (Fig.~\ref{fig:pbnet}) mirrors the basic structure of Alg.~\ref{alg:pgd}. In implementation, it requires all learnable parameters as input (measurement matrix, step size, and sparsity penalty), initialization, ground truth, and number of iterations (or depth of network). As output, the network returns the final estimate of the sparse recovery optimization.

The specific Pytorch flags and functionalities that enable us to properly use the automatic differentiator for sparse recovery and for physics-based learning are contained in lines 14, 15, and 26. Lines 14 and 15 set up the initialization to the network, first in line 14 detaching it from any previous automatic differentiation graphs and second in line 15 setting its \textit{requires$\_$grad} field to be {\color{green}True}. This will allow the automatic differentiation to track operations related to $\x$ for taking derivatives of $\D(\x;\y)$ with respect to $\x$. Line 26 sets the \textit{create$\_$graph} flag of the automatic differentiator to {\color{green}True}. This tells Pytorch to store these operations so that they can be traced through by a second automatic differentiator for computing derivatives with respect to learnable parameters at training time.

While some physics-based networks take in measurements as input, this is often not possible when learning a system's experimental design. Here, in line 17, the measurements are synthesized using the learnable measurement matrix and ground truth sparse vector. 

\subsection*{Physics-based Learning}

The mechanics of training a physics-based network are similar to that of training any neural net or convolutional neural network in that it relies on a dataset, an optimizer, and automatic differentiation to compute gradients. In our particular application, our dataset consists of sparse vectors with amplitudes that are normally distributed. As for the optimizer, we use the adaptive moment estimation (Adam)~\cite{kingma2014adam} algorithm to perform the gradient updates. For researchers that interested in learning for large-scale computational imaging systems, automatic differentiation will require more memory than available on commercial graphical processing units. To enable learning at these scales, vanilla automatic differentiation can be replaced with memory-efficient techniques~\cite{kellman2020memory}.

\section{Remarks}

In this tutorial we overview the implementation of physics-based learning in Pytorch with a simple open-source implementation that provides a minimum working example for those looking to optimize their own system. The concepts discussed here generalize to vast sets of applications and optimization algorithms.

\bibliographystyle{icml2019}
\bibliography{references}

\begin{thebibliography}{11}
\providecommand{\natexlab}[1]{#1}
\providecommand{\url}[1]{\texttt{#1}}
\expandafter\ifx\csname urlstyle\endcsname\relax
  \providecommand{\doi}[1]{doi: #1}\else
  \providecommand{\doi}{doi: \begingroup \urlstyle{rm}\Url}\fi

\bibitem[Boyd et~al.(2011)Boyd, Parikh, Chu, Peleato, Eckstein,
  et~al.]{boyd2011distributed}
Boyd, S., Parikh, N., Chu, E., Peleato, B., Eckstein, J., et~al.
\newblock Distributed optimization and statistical learning via the alternating
  direction method of multipliers.
\newblock \emph{Foundations and Trends{\textregistered} in Machine learning},
  3\penalty0 (1):\penalty0 1--122, 2011.

\bibitem[Geman \& Yang(1995)Geman and Yang]{geman1995nonlinear}
Geman, D. and Yang, C.
\newblock Nonlinear image recovery with half-quadratic regularization.
\newblock \emph{IEEE transactions on Image Processing}, 4\penalty0
  (7):\penalty0 932--946, 1995.

\bibitem[Gregor \& LeCun(2010)Gregor and LeCun]{Gregor:2010}
Gregor, K. and LeCun, Y.
\newblock Learning fast approximations of sparse coding.
\newblock In \emph{Proceedings of the 27th International Conference on
  International Conference on Machine Learning}, ICML'10, pp.\  399--406, 2010.

\bibitem[Griewank \& Walther(2008)Griewank and Walther]{Griewank2018evalder}
Griewank, A. and Walther, A.
\newblock \emph{Evaluating Derivatives: Principles and Techniques of
  Algorithmic Differentiation}.
\newblock Society for Industrial and Applied Mathematics, Philadelphia, PA,
  USA, second edition, 2008.
\newblock ISBN 0898716594, 9780898716597.

\bibitem[Kellman et~al.(2019)Kellman, Bostan, Repina, and
  Waller]{kellman2019physics}
Kellman, M., Bostan, E., Repina, N., and Waller, L.
\newblock Physics-based learned design: Optimized coded-illumination for
  quantitative phase imaging.
\newblock \emph{IEEE Transactions on Computational Imaging}, 5\penalty0
  (3):\penalty0 344--353, 2019.

\bibitem[Kellman et~al.(2020)Kellman, Zhang, Tamir, Bostan, Lustig, and
  Waller]{kellman2020memory}
Kellman, M., Zhang, K., Tamir, J., Bostan, E., Lustig, M., and Waller, L.
\newblock Memory-efficient learning for large-scale computational imaging.
\newblock \emph{arXiv preprint arXiv:2003.05551}, 2020.

\bibitem[Kingma \& Ba(2014)Kingma and Ba]{kingma2014adam}
Kingma, D.~P. and Ba, J.
\newblock Adam: A method for stochastic optimization.
\newblock \emph{arXiv preprint arXiv:1412.6980}, 2014.

\bibitem[Ongie et~al.(2020)Ongie, Jalal, Metzler, Baraniuk, Dimakis, and
  Willett]{ongie2020deep}
Ongie, G., Jalal, A., Metzler, C.~A., Baraniuk, R.~G., Dimakis, A.~G., and
  Willett, R.
\newblock Deep learning techniques for inverse problems in imaging.
\newblock \emph{arXiv preprint arXiv:2005.06001}, 2020.

\bibitem[Parikh \& Boyd(2014)Parikh and Boyd]{parikh2014proximal}
Parikh, N. and Boyd, S.
\newblock Proximal algorithms.
\newblock \emph{Foundations and Trends{\textregistered} in Optimization},
  1\penalty0 (3):\penalty0 127--239, 2014.

\bibitem[Paszke et~al.(2017)Paszke, Gross, Chintala, Chanan, Yang, DeVito, Lin,
  Desmaison, Antiga, and Lerer]{paszke2017automatic}
Paszke, A., Gross, S., Chintala, S., Chanan, G., Yang, E., DeVito, Z., Lin, Z.,
  Desmaison, A., Antiga, L., and Lerer, A.
\newblock Automatic differentiation in pytorch.
\newblock 2017.

\bibitem[Sun et~al.(2016)Sun, Li, Xu, et~al.]{sun2016deep}
Sun, J., Li, H., Xu, Z., et~al.
\newblock Deep {ADMM-Net} for compressive sensing {MRI}.
\newblock In \emph{Advances in Neural Information Processing Systems}, pp.\
  10--18, 2016.

\end{thebibliography}
\end{document}